\documentclass[aps,pre,preprint]{revtex4-2}

\usepackage{graphicx}

\usepackage{amsmath} %
\usepackage{etoolbox}

\newcommand{\I}{\mathrm{j}}

\newcommand{\E}{\mathrm{e}}
\newcommand{\D}{\mathrm{d}}

\newcommand{\graphicsfolder}{.}

\begin{document}


\title{Simple Circuit Equivalents for the Constant Phase Element}


\author{Sverre Holm}
\affiliation{Department of Physics, University of Oslo, Norway}

\author{Thomas Holm}
\affiliation{Clean Energy Research Centre, The University of British Columbia, Vancouver, Canada}

\author{{\O}rjan Gr{\o}ttem Martinsen}
\affiliation{Department of Physics, University of Oslo, Norway}
\affiliation{Department of Clinical and Biomedical Engineering, Oslo University Hospital, Norway}


\date{\today}

\begin{abstract}
The constant phase element (CPE) with a frequency-independent negative phase between current and voltage is used extensively in e.g.~the bioimpedance and electrochemistry fields. Its physical meaning is only partially understood. Here we show that the responses of both the  common capacitive CPE as well as the inductive CPE are exactly the same as those of simple RL and RC circuits where the inductor's or capacitor's value increases linearly with time. The resulting step and impulse responses are found and verified by simulation with the Micro-Cap simulation program. The realization with time-varying components correlates with known time-varying properties in applications. 
\end{abstract}


\maketitle


\section{Introduction}

\subsection{Constant phase element }

The constant phase element (CPE) is a capacitive impedance with a phase angle in the range $\langle -\pi/2, 0 \rangle$ which is independent of  frequency. 
It was first introduced by Cole in  connection with the electrical impedance of suspensions of spheres \cite{cole1928electric} and of cell membranes \cite{cole1940permeability}. Jonscher observed that this model is valid for a large range of dielectrics, calling it the ``universal'' dielectric response. He also connected it to the temporal power law response of the Curie-von-Schweidler law \cite{jonscher1977universal} which was first observed for real-life capacitors more than a century ago. Further Westerlund observed that the response function  may be expressed with a non-integer, fractional, derivative \cite{westerlund1991dead} and an effective time-varying capacitance with a  power-law time variation. In \cite{fouda2020nonlinear} it was found that the CPE could be modeled as a voltage-dependent capacitance which makes it into a nonlinear circuit.

The constant phase element is applied for modeling experimental data and here in particular the bioimpedance and electrochemical impedance fields will be highlighted. In the first field, the CPE is  a common model \cite[Ch. 9.2.5]{grimnes201bioimpedance} along with the related Cole impedance model  \cite[Ch ~5.8]{holm2019waves}. Often it is interpreted as a distribution of time constants due to a statistical distribution of cell sizes resulting in a band-limited approximation to a CPE as demonstrated in \cite[Ch.~7.2]{holm2019waves}. 

In electrochemistry, CPE behavior is also commonly observed experimentally, and has been interpreted as a statistical distribution of time constants due to for example crystal orientation, surface roughness, and resistance distribution in an oxide layer at the surface  \cite[Ch.~8]{lasia2002electrochemical}, \cite[Ch.~13]{orazem2008electrochemical}, and \cite{hirschorn2010constant}. While several physical explanations exist for CPE behavior, such behavior is observed even in idealized experiments  \cite{sharma2019experimental} and equivalent circuit fitting using CPE elements  is often used without prior justification. The CPE model is then often used with resistors in series and parallel. That is also the case in bioimpedance \cite{martinsen1997dielectric} as well as in the modeling of supercapacitors \cite{zhang2018review}.

Here we show that CPE behavior can result from simple circuits with time-varying component values. The common capacitive CPE with a negative phase angle will have the same current response to an input voltage impulse or step as a resistor in series with an inductor that increases linearly with time. An inductive  CPE with a positive phase angle will have  the same voltage response as a resistor in parallel with a linearly increasing capacitor. The models are inspired by similar ones in linear viscoelasticity \cite{pandey2016linking, yang2020novel} where they may model viscosity due to a stick-slip motion between grains in a water-saturated sediment  
 \cite{buckingham2000wave}.

The paper starts with defining the CPE and connects its frequency and time responses with the fractional derivative description.  The  proposed linearly time-varying circuits are then analyzed analytically and both exact and approximate solutions are found. The results are confirmed by simulation with Micro-Cap 12.

\section{Models}

Here, tilde is used to denote a Fourier transform so a function $f(t)$ has a Fourier transform $\tilde{f}(\omega)$. 
%
\subsection{Capacitive Constant Phase Element}

The capacitive constant phase element has an impedance given by \cite{westerlund1991dead,  fouda2020nonlinear}:
%
\begin{equation}
\tilde{Z}(\omega) =\frac{\tilde{u}(\omega)}{\tilde{i}(\omega)}= \frac{1}{(\I \omega)^{\alpha} C_{\alpha}}, \enspace 0<\alpha \le 1.
\label{eq:fractionalC}
\end{equation}
where $C_\alpha$ is the CPE parameter in  $\mathrm{F} \cdot \mathrm{s}^{\alpha-1}$ and $\alpha$ is the order.  
The impedance has a negative phase angle of $-\alpha \pi/2$ which varies from $0$ to $-\pi/2$. The value $\alpha=1$ gives an ordinary capacitor, $a = 0$ is a resistor, and $a = 0.5$ corresponds to the Warburg element used to model diffusion processes. The CPE model corresponds to an equivalent complex relative permittivity 
\begin{equation}
\tilde{\varepsilon}(\omega) = \frac{C_\alpha d}{\varepsilon_0 A (\I \omega)^{1-\alpha}},
\end{equation}
where $d$ and $A$ are the equivalent plate distance and plate area of the capacitor respectively and $\varepsilon_0$ is the permittivity of vacuum.
%
%
The inverse Fourier transform of \eqref{eq:fractionalC} is a convolution of the input voltage with the impulse response which here  is a temporal power law  \cite[App.~A.3]{holm2019waves}: 
\begin{equation}
i_{imp}(t) = \frac{C_\alpha}{\Gamma(-\alpha)} t^{-\alpha-1}. \enspace t>0.
\label{eq:impulseResp}
\end{equation}
The step response is the integral of the impulse response and is also a power law function:
\begin{equation}
i_{step}(t) = \frac{-C_\alpha}{\Gamma(-\alpha)\alpha} t^{-\alpha} 
\label{eq:stepResp}
\end{equation}
This is the Curie-von-Schweidler law  \cite{jonscher1977universal, westerlund1991dead}. Both responses have an initial singularity which indicates that the CPE model is a simplified and idealized model of a real-life phenomenon.

As a power law in the frequency domain is one way of defining a fractional derivative, the CPE of \eqref{eq:fractionalC} is also equivalent to a fractional capacitor, \cite{westerlund1991dead}, \cite[Ch.~5.8]{holm2019waves}:
\begin{equation}
i(t) = C_{\alpha} \frac{\partial^\alpha u(t)}{\partial t^\alpha}.
\end{equation}

\subsection{Inductive Constant Phase Element}
An inductive constant phase element where the impedance has a positive phase angle $\alpha \pi/2$, can be described by 
\begin{equation}
\tilde{Z}(\omega) = (\I \omega)^{\alpha} L_{\alpha},  \enspace 0<\alpha \le 1,
\label{eq:fractionalL}
\end{equation}
where $L_\alpha$ is in units of $\mathrm{H} \cdot \mathrm{s}^{\alpha-1}$  and $\alpha=1$ gives an ordinary inductance. The voltage is a fractional derivative of the current:
\begin{equation}
u(t) = L_{\alpha} \frac{\partial^\alpha i(t)}{\partial t^\alpha}.
\end{equation}
Both the capacitive and the inductive CPEs are included in the definition of the fractance found for instance in \cite{radwan2012expression, abdelouahab2014memfractance}.


\section{Circuit realization of the Constant Phase Element}

We model the CPE using time-varying inductors and capacitors. They don't necessarily exist as physical devices although some devices may approximate them. Building on circuit theory where the usual definition is that  magnetic flux is the product of inductance and current, the current-voltage characteristics is:
\begin{equation}
u(t) = \frac{\D \Phi(t)}{\D t} = \frac{\D \left[ L(t) \cdot i(t)\right]}{\D t} =  L(t) \cdot \frac{\D i(t)}{\D t} + \frac{\D L(t)}{\D t}i(t),
\label{eq:voltageFlux}
\end{equation}
where   $L(t)$ is a time-varying inductance. In a time-varying capacitor current will have a similar  relationship with charge.
%
This is different from the field of linear viscoelasticity which has inspired this paper. There the relationship is defined by the first term only \cite{Chhabra2010}:
\begin{equation}
\sigma(t) = \eta(t) \cdot \frac{\D \varepsilon(t)}{\D t},
\end{equation}
where $\sigma$ is stress, $\varepsilon$ is strain and $\eta(t)$ is the apparent time-varying viscosity as used in \cite{buckingham2000wave, pandey2016linking, yang2020novel}.  

This ambiguity is reflected in how time-varying inductors and capacitors are implemented  in circuit simulators. Micro-Cap from Spectrum Software implements both terms of \eqref{eq:voltageFlux}. On the other hand OrCAD PSpice only implements the first term and this is seen as a problem in \cite{biolek2007modeling}. Simulink\textsuperscript{\tiny\textregistered} from The Mathworks, Inc.~gives the user a choice of  whether to include the last term or not. Here both terms of \eqref{eq:voltageFlux} will be used, but it will turn out that since the variation in inductance or capacitive is linear with time, the omission of the last term will not change the final result much.

\subsection{Capacitive Constant Phase Element}

\begin{figure}[h]
	\centering
	\begin{minipage}[b]{0.4\textwidth}
		\includegraphics{\graphicsfolder/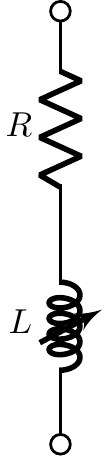}%
	\end{minipage}
	\hfill
  	\begin{minipage}[b]{0.4\textwidth}
  		\includegraphics{\graphicsfolder/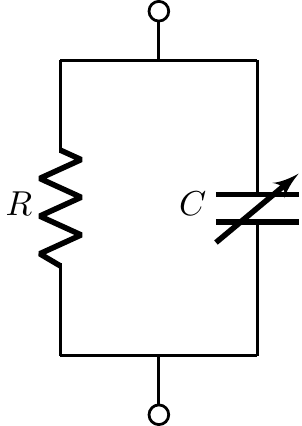}%
	\end{minipage}
	\caption{Equivalent circuits for a capacitive (left) and an inductive (right) Constant Phase Element when inductance and capacitance increase linearly with time} 
	\label{fig:Circuit-CPEs}
\end{figure}

The left-hand circuit of Fig.~\ref{fig:Circuit-CPEs} where $L(t)= L_0 + \theta t_2$ is an inductance that increases linearly with time. It has an instantaneous impedance of: 
\begin{equation}
\tilde{Z}(\I \omega,t_2) = \left( R + \theta \right) + \I \omega \left( L_0 + \theta t_2 \right),
\label{eq:impedanceCPE-C}
\end{equation}
where $\theta$ has unit H/s = $\Omega$. The only effect of including the second term of \eqref{eq:voltageFlux} in the definition of time-varying inductance is that $\theta$ contributes to the effective resistance. 

A second time variable, $t_2$ is used here as the time of the input impulse, $t$, and the time when the inductance starts changing, $t_2$, in general are  independent of each other. We will however assume that the variable inductor starts changing at the exact moment when the input impulse is applied so $t_2 = t$. The voltage-current relationship is then:
\begin{equation}
u(t) = Z(t)i(t) = R i +  \frac{\D \left[(L_0 + \theta t)i(t)  \right]}{\D t} = (R+\theta ) i + \left( L_0 +\theta t \right) \frac{\D i}{\D t} 
\end{equation}
By applying a voltage impulse as input at time $t=0$,  the current will follow:
\begin{equation}
\frac{{\D i}/{\D t}}{i} =  - \frac{R+\theta }{L_0 + \theta t}, \enspace t>0
\label{eq:CPE-pde}
\end{equation}
Following \cite[App.~1]{pandey2016linking}, integration of \eqref{eq:CPE-pde} gives
\begin{equation}
\ln i = -\frac{R+\theta}{\theta}\left\{ \ln{(L_0 + \theta t)} + \ln K \right\}
\label{eq:lnRelation}
\end{equation}
where $K$ is a constant determined by the initial conditions.

\subsubsection{Exact CPE}
Assume now that the initial value of the inductance, $L_0$, is zero. Taking the exponential of the previous equation then gives:
\begin{equation}
i(t) \propto \left(\theta t \right)^{-(R+\theta)/\theta}, \enspace t>0.
\label{eq:exactCPE2}
\end{equation}
The constant of proportionality can be determined by considering a practical case related to the following simulation in Micro-Cap. If the input unit impulse is implemented by exciting one sample of duration $t_s$  with a voltage of $1/t_s$ then the initial condition is that $i(t_s)= 1/(R t_s)$. Thus the  final result will be
\begin{equation}
i(t) =  \frac{t_s^\alpha}{R} t^{-\alpha-1}, \enspace \alpha=\frac{R}{\theta}, \enspace t>0.
\label{eq:exactCPE3}
\end{equation}
Surprisingly this simple circuit has exactly the same response as the CPE of \eqref{eq:impulseResp} for all positive values of time. 

\subsubsection{Approximate CPE}
As it may be hard to imagine a practical time-varying circuit where a component value starts with value zero, we will now let $L_0>0$. Then  the solution is:
\begin{equation}
i(t) = \frac{1}{R} \left(1 + \frac{\theta}{L_0} t \right)^{-(R+\theta)/\theta}, \enspace t>0.
\label{eq:exactCPE1}
\end{equation}
Assuming that $t \gg \tau$ so that the second term in the parenthesis dominates this is
\begin{equation}
i(t) \approx \frac{1}{R} \left( \frac{t}{\tau} \right)^{-\alpha-1}, \enspace \tau = \frac{L_0}{\theta}, \enspace \alpha=\frac{R}{\theta}, \enspace t>0.
\label{eq:CPE1-time}
\end{equation}
This is also equivalent to \eqref{eq:impulseResp} and therefore this time-varying circuit approximates a CPE.

\subsubsection {Remarks}
The  results of \eqref{eq:exactCPE3} and \eqref{eq:CPE1-time} demonstrate that the time-varying circuit on the left-hand side of  Fig.~\ref{fig:Circuit-CPEs} has the same current response to an input voltage as the CPE. It should be noted that due to the time-variation in the circuit, this does not imply that the opposite is true, i.e.~that the voltage response to an input current is the same as for a CPE. This is a limitation of the model. Despite being time-varying, the system is linear and therefore the step response can be found by integrating the impulse responses just as when going from \eqref{eq:impulseResp} to \eqref{eq:stepResp}.


\subsection{Inductive Constant Phase Element}

The right-hand circuit of Fig.~\ref{fig:Circuit-CPEs}  is similar to the left-hand one in the sense that the admittance follows the expression for the impedance  of \eqref{eq:impedanceCPE-C}:
\begin{equation}
\tilde{Y}(\I \omega,t_2) = \left( \frac{1}{R} +\theta \right) + \I \omega \left(C_0 + \theta t_2 \right)
\label{eq:admittanceCPE-L}
\end{equation}
where the time-varying component is now a capacitor and $\theta$ has unit F/s = S. 
Therefore when $C_0=0$ the response will parallel that of \eqref{eq:exactCPE3} and be:
\begin{equation}
u(t) =  R \tau_0^\alpha t^{-\alpha-1}, \enspace  \alpha=\frac{1}{R\theta}, \enspace t>0.
\end{equation}
Further, it will have a voltage response to an impulse in current which will be of the same form as \eqref{eq:CPE1-time} when $C_0>0$: 
\begin{equation}
u(t) \approx R \left( \frac{t}{\tau} \right)^{-\alpha-1}, \enspace \tau = \frac{C_0}{\theta}, \enspace  \alpha=\frac{1}{R\theta},  \enspace t>0.
\end{equation}

%

\section{Verification by simulation}

\subsection{Constant Phase Element}
\begin{figure}[t]
	\includegraphics[width=0.9\columnwidth]{\graphicsfolder/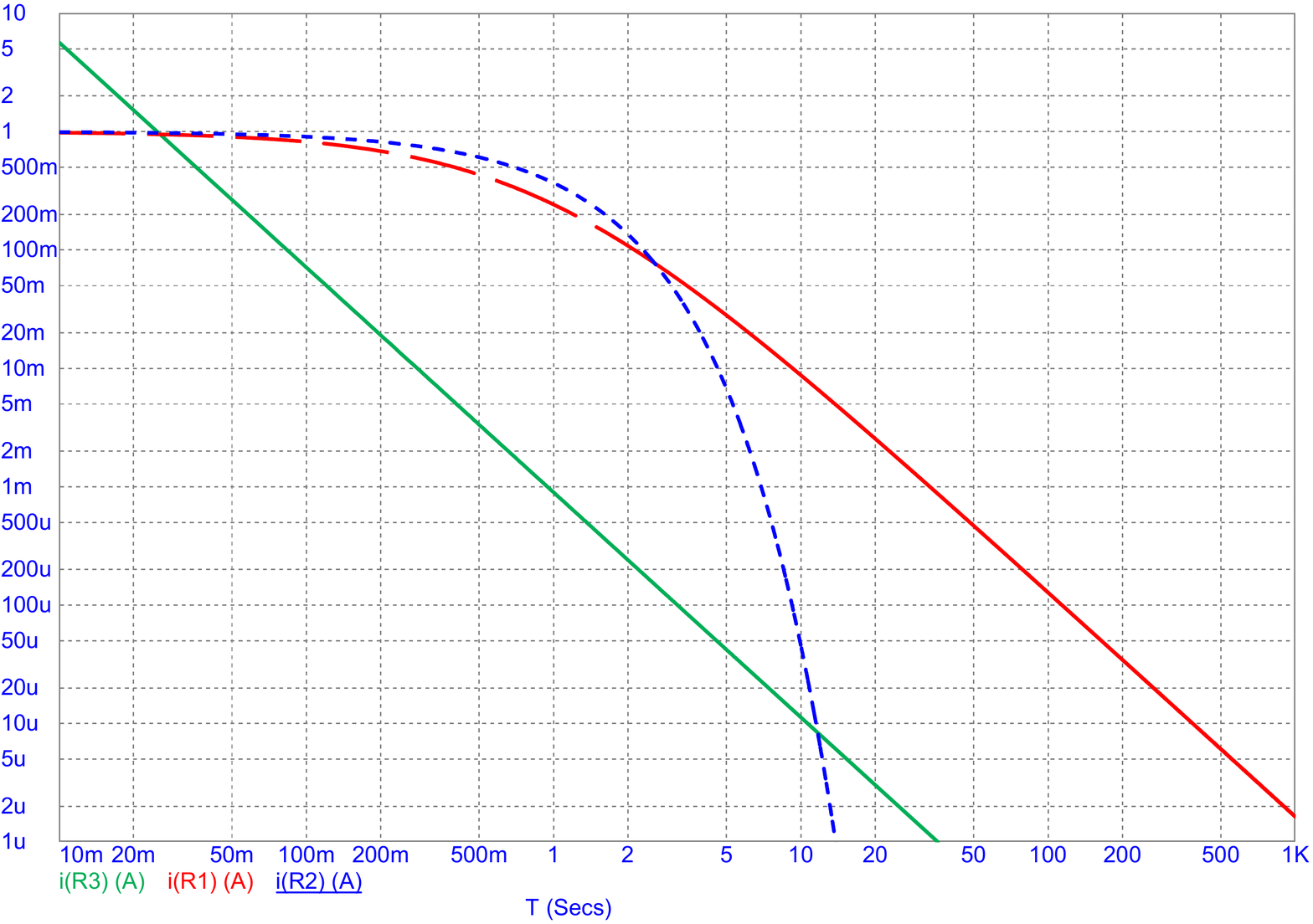}
	\caption{Micro-Cap 12 simulation of current response to a voltage impulse  for the left-hand circuit of  Fig.~\ref{fig:Circuit-CPEs} with $R=1$, $L=1+t/0.9$, i.e.~$\tau = \alpha=0.9$ (red), $R=1$, $L=t/0.9$, i.e.~$\alpha=0.9$ (green) compared to an ordinary RL-circuit with response $\E^{-t/\tau}/R$, \, $\tau = L/R, \, L=1, R=1$ (blue, dashed)}
	\label{fig:MicroCapLogCPE1-0.9}
\end{figure}

%
Micro-Cap version 12 \cite{SpectrumSoftware} is a versatile tool for simulating complex circuits. It is used for the normalized case with $R=1$, $L=1+t/0.9$ and thus $\theta=\alpha=0.9$ as shown in Fig.~\ref{fig:MicroCapLogCPE1-0.9}. In this figure as well as the two next ones, a comparison is also made with  a time-invariant circuit with $R=1$, $L=1$ with an exponential response.

\begin{figure}[t]
	\includegraphics[width=0.9\columnwidth]{\graphicsfolder/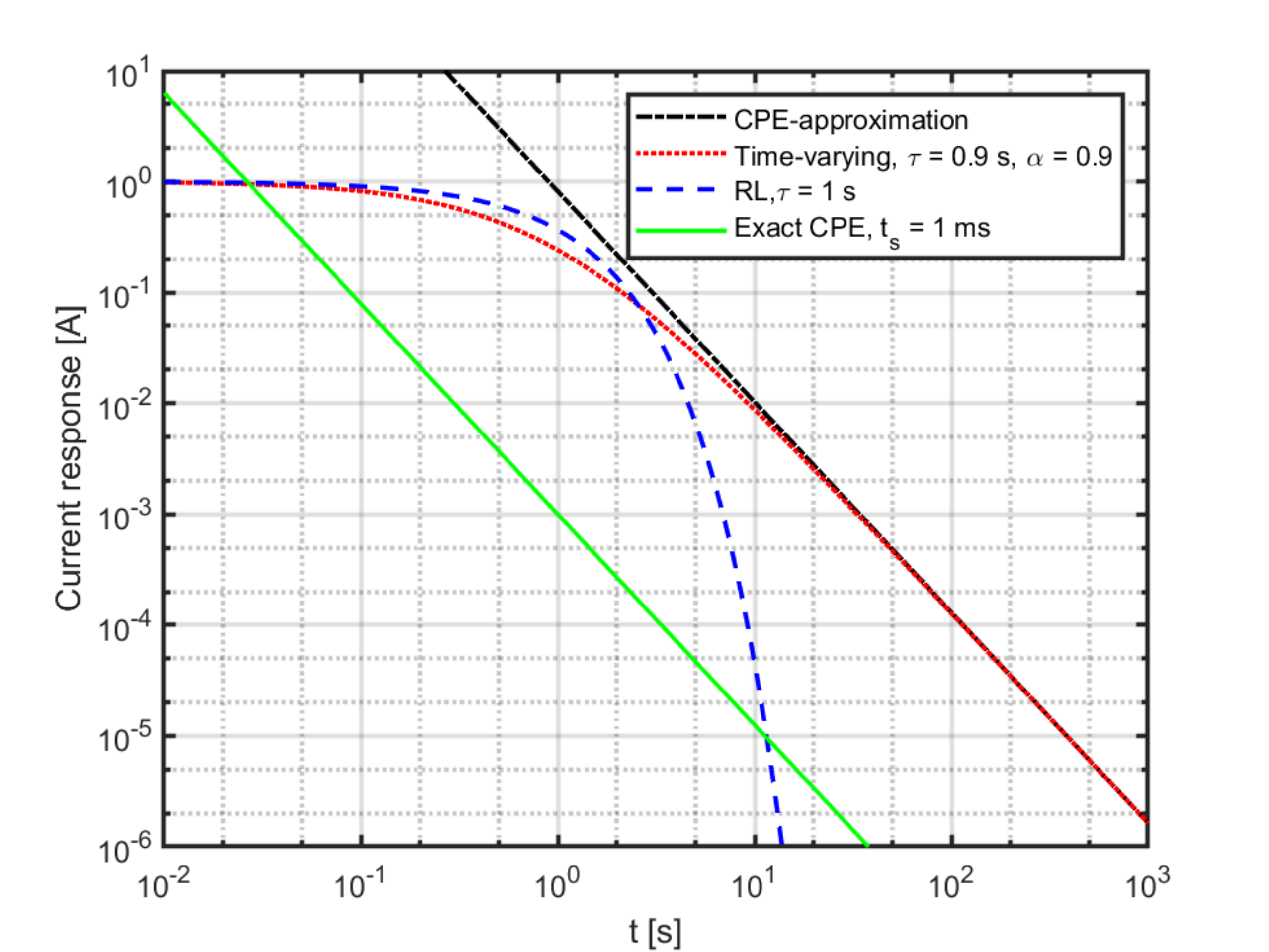}%
	\caption{Current response to a voltage impulse  for  the left-hand circuit of  Fig.~\ref{fig:Circuit-CPEs} for $\alpha=0.9$ computed in Matlab. Grid lines are the same as in Fig.~\ref{fig:MicroCapLogCPE1-0.9}}
	\label{fig:LogCPE1-0.9}
\end{figure}

\begin{figure}[t]
	\includegraphics[width=0.9\columnwidth]{\graphicsfolder/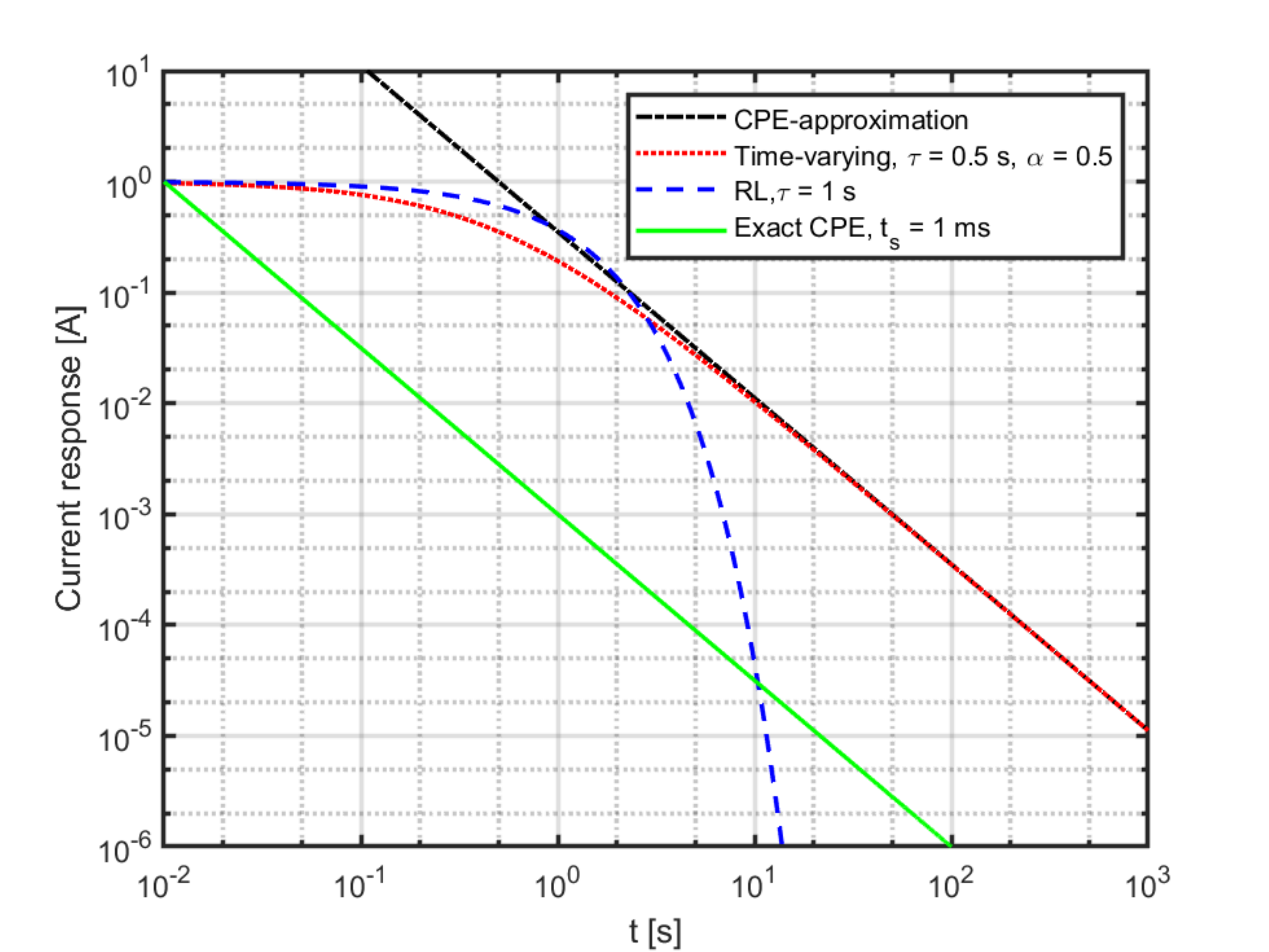}%
	\caption{Current response to a voltage impulse  for  the left-hand circuit of  Fig.~\ref{fig:Circuit-CPEs} for $\alpha=0.5$ (Warburg element) computed in Matlab}
	\label{fig:LogCPE1-0.5}
\end{figure}

Equations \eqref{eq:exactCPE1} and \eqref{eq:CPE1-time} were then implemented in Matlab and plotted on a logarithmic grid. 
The case for $\alpha=0.9$ is shown in Fig.~\ref{fig:LogCPE1-0.9}. 
The discrepancy between the Matlab computation and the Micro-Cap simulation is maximally in the order of 0.1\%.
The case for $R=1$, $L=1+ 0.5 t$ and thus $\theta=\alpha=0.5$ corresponding to a Warburg element, is shown in Fig.~\ref{fig:LogCPE1-0.5}.

The exact expression and the CPE approximation are similar for  time above 4-5 seconds when $\tau=0.9$, as shown in Fig.~\ref{fig:LogCPE1-0.9}. 
In the case of $\tau=0.5$, Fig.~\ref{fig:LogCPE1-0.5}, 
 it happens after a few seconds. This is as expected from the theory  as the approximation of \eqref{eq:CPE1-time} is valid when $t>>\tau$.
 
 \section{Discussion}
 

%
%

Modeling of the capacitive and inductive constant phase elements with time-varying circuits does not critically depend on whether the second term of \eqref{eq:voltageFlux} is included in the definition of a time-varying inductor or not.  The second term only adds $\theta$ to R in the results. Thus \eqref{eq:exactCPE2} and \eqref{eq:exactCPE1} has $(R+\theta)/\theta$ rather than $R/\theta$ in the exponents. The functional form of the final result is therefore independent of the definition, but it will change the exponents by 1.

There are several examples of  time dependent electrical parameters in biological materials justifying the modeling of the CPE with a time-varying circuit. One candidate is the memristance displayed for example by human skin \cite{johnsen2011memristive, pabst2018non}. The resistance changes as a function of the net amount of charge having passed through the material and when the electrical current is reversed, the resistance will change in the opposite direction. Although the change is charge driven, it will appear as a time dependent resistor when applying a periodic AC signal.  
Similar mechanisms have been reported for capacitance and inductance, named memcapacitance and meminductance \cite{di2009circuit} and unpublished results from our group indicate that some biomaterials also have memcapacitive properties. 
Furthermore, bioimpedance measurements will often contain elements of non-linear properties, typically when using small electrodes where the current density in some volumes may exceed the linear range \cite{yamamoto1981non, kalvoy2009impedance}. 

Another example of a system that resets itself for every polarity change is the charging of a double-layer in an electrochemical system due to an imposed signal. In each period, the applied signal leads to movements of ions in the solution towards or away from the electrode surface. During charging, ions move towards the electrode, and give electrostatic resistance to subsequent movement of ions. Similarly, during discharge, the ions move away from the electrode and give resistance to the discharge process which varies with time.

These properties open the possibility for finding the equivalent of a steady-state transfer response. The challenge is to justify elimination of the second temporal variable $t_2$ in \eqref{eq:impedanceCPE-C} as the Fourier relation between transfer function and impulse response strictly speaking no longer holds for a time-varying system as also discussed in \cite{pandey2016linking}. If the second temporal variable is reset at every zero-crossing, it will be easier to justify an approximate Fourier relation between the impulse response and the transfer function. %





Finally, a goal is often to translate the non-ideal CPE element into an equivalent capacitor, notable examples being the Brug equation \cite{brug1984analysis}, the oxide layer model   \cite{hirschorn2010determination}, and  the approaches of  \cite{hsu2001concerning, allagui2016reevaluation, gharbi2020revisiting}. The interpretation is often limited to specific systems, where the Brug equation is the most widely applied. 
While limited to a voltage input pulse here, the presented equivalence circuits may contribute to understanding overall CPE behavior and subsequently establishing more robust methods of interpretation.

%

\section{Conclusion}

We have shown that the capacitive constant phase element (CPE) has exactly the same current response as a resistor in series with a linearly increasing inductance. Likewise the inductive CPE has a voltage response which is similar to that of a resistor in parallel with a linearly increasing capacitance. These forms of  CPE are both subsets of what has recently been called a fractance. The Micro-Cap 12 simulation program is used to verify that the impulse responses follow the temporal power law predicted by theory. The realization with time-varying components correlates with known time-varying properties in applications, but this property will need to be explored and established more firmly in future work.

\vspace{1 cm}

\begin{acknowledgments}
The authors would like to thank Dr.~Vikash Pandey for helpful discussions of step vs impulse responses.
\end{acknowledgments}

%


\begin{thebibliography}{31}%
\makeatletter
\providecommand \@ifxundefined [1]{%
 \@ifx{#1\undefined}
}%
\providecommand \@ifnum [1]{%
 \ifnum #1\expandafter \@firstoftwo
 \else \expandafter \@secondoftwo
 \fi
}%
\providecommand \@ifx [1]{%
 \ifx #1\expandafter \@firstoftwo
 \else \expandafter \@secondoftwo
 \fi
}%
\providecommand \natexlab [1]{#1}%
\providecommand \enquote  [1]{``#1''}%
\providecommand \bibnamefont  [1]{#1}%
\providecommand \bibfnamefont [1]{#1}%
\providecommand \citenamefont [1]{#1}%
\providecommand \href@noop [0]{\@secondoftwo}%
\providecommand \href [0]{\begingroup \@sanitize@url \@href}%
\providecommand \@href[1]{\@@startlink{#1}\@@href}%
\providecommand \@@href[1]{\endgroup#1\@@endlink}%
\providecommand \@sanitize@url [0]{\catcode `\\12\catcode `\$12\catcode
  `\&12\catcode `\#12\catcode `\^12\catcode `\_12\catcode `\%12\relax}%
\providecommand \@@startlink[1]{}%
\providecommand \@@endlink[0]{}%
\providecommand \url  [0]{\begingroup\@sanitize@url \@url }%
\providecommand \@url [1]{\endgroup\@href {#1}{\urlprefix }}%
\providecommand \urlprefix  [0]{URL }%
\providecommand \Eprint [0]{\href }%
\providecommand \doibase [0]{https://doi.org/}%
\providecommand \selectlanguage [0]{\@gobble}%
\providecommand \bibinfo  [0]{\@secondoftwo}%
\providecommand \bibfield  [0]{\@secondoftwo}%
\providecommand \translation [1]{[#1]}%
\providecommand \BibitemOpen [0]{}%
\providecommand \bibitemStop [0]{}%
\providecommand \bibitemNoStop [0]{.\EOS\space}%
\providecommand \EOS [0]{\spacefactor3000\relax}%
\providecommand \BibitemShut  [1]{\csname bibitem#1\endcsname}%
\let\auto@bib@innerbib\@empty
\bibitem [{\citenamefont {Cole}(1928)}]{cole1928electric}%
  \BibitemOpen
  \bibfield  {author} {\bibinfo {author} {\bibfnamefont {K.~S.}\ \bibnamefont
  {Cole}},\ }\bibfield  {title} {\bibinfo {title} {Electric impedance of
  suspensions of spheres},\ }\href@noop {} {\bibfield  {journal} {\bibinfo
  {journal} {J. Gen. Physiol.}\ }\textbf {\bibinfo {volume} {12}},\ \bibinfo
  {pages} {29} (\bibinfo {year} {1928})}\BibitemShut {NoStop}%
\bibitem [{\citenamefont {Cole}(1940)}]{cole1940permeability}%
  \BibitemOpen
  \bibfield  {author} {\bibinfo {author} {\bibfnamefont {K.~S.}\ \bibnamefont
  {Cole}},\ }\bibfield  {title} {\bibinfo {title} {Permeability and
  impermeability of cell membranes for ions},\ }in\ \href@noop {} {\emph
  {\bibinfo {booktitle} {Cold Spring Harbor Symposia on Quantitative
  Biology}}},\ Vol.~\bibinfo {volume} {8}\ (\bibinfo {organization} {Cold
  Spring Harbor Laboratory Press},\ \bibinfo {year} {1940})\ pp.\ \bibinfo
  {pages} {110--122}\BibitemShut {NoStop}%
\bibitem [{\citenamefont {Jonscher}(1977)}]{jonscher1977universal}%
  \BibitemOpen
  \bibfield  {author} {\bibinfo {author} {\bibfnamefont {A.~K.}\ \bibnamefont
  {Jonscher}},\ }\bibfield  {title} {\bibinfo {title} {The ‘universal’
  dielectric response},\ }\href@noop {} {\bibfield  {journal} {\bibinfo
  {journal} {Nature}\ }\textbf {\bibinfo {volume} {267}},\ \bibinfo {pages}
  {673} (\bibinfo {year} {1977})}\BibitemShut {NoStop}%
\bibitem [{\citenamefont {Westerlund}(1991)}]{westerlund1991dead}%
  \BibitemOpen
  \bibfield  {author} {\bibinfo {author} {\bibfnamefont {S.}~\bibnamefont
  {Westerlund}},\ }\bibfield  {title} {\bibinfo {title} {Dead matter has
  memory!},\ }\href@noop {} {\bibfield  {journal} {\bibinfo  {journal} {Phys.
  Scr.}\ }\textbf {\bibinfo {volume} {43}},\ \bibinfo {pages} {174} (\bibinfo
  {year} {1991})}\BibitemShut {NoStop}%
\bibitem [{\citenamefont {Fouda}\ \emph {et~al.}(2020)\citenamefont {Fouda},
  \citenamefont {Allagui}, \citenamefont {Elwakil}, \citenamefont {Das},
  \citenamefont {Psychalinos},\ and\ \citenamefont
  {Radwan}}]{fouda2020nonlinear}%
  \BibitemOpen
  \bibfield  {author} {\bibinfo {author} {\bibfnamefont {M.~E.}\ \bibnamefont
  {Fouda}}, \bibinfo {author} {\bibfnamefont {A.}~\bibnamefont {Allagui}},
  \bibinfo {author} {\bibfnamefont {A.~S.}\ \bibnamefont {Elwakil}}, \bibinfo
  {author} {\bibfnamefont {S.}~\bibnamefont {Das}}, \bibinfo {author}
  {\bibfnamefont {C.}~\bibnamefont {Psychalinos}},\ and\ \bibinfo {author}
  {\bibfnamefont {A.~G.}\ \bibnamefont {Radwan}},\ }\bibfield  {title}
  {\bibinfo {title} {Nonlinear charge-voltage relationship in constant phase
  element},\ }\href@noop {} {\bibfield  {journal} {\bibinfo  {journal}
  {AEU-Int. J. Electron. .C}\ }\textbf {\bibinfo {volume} {117}},\ \bibinfo
  {pages} {153104} (\bibinfo {year} {2020})}\BibitemShut {NoStop}%
\bibitem [{\citenamefont {Grimnes}\ and\ \citenamefont
  {Martinsen}(2015)}]{grimnes201bioimpedance}%
  \BibitemOpen
  \bibfield  {author} {\bibinfo {author} {\bibfnamefont {S.}~\bibnamefont
  {Grimnes}}\ and\ \bibinfo {author} {\bibfnamefont {{\O}.~G.}\ \bibnamefont
  {Martinsen}},\ }\href@noop {} {\emph {\bibinfo {title} {Bioimpedance and
  Bioelectricity Basics;}}}\ (\bibinfo  {publisher} {Academic Press},\ \bibinfo
  {year} {2015})\BibitemShut {NoStop}%
\bibitem [{\citenamefont {Holm}(2019)}]{holm2019waves}%
  \BibitemOpen
  \bibfield  {author} {\bibinfo {author} {\bibfnamefont {S.}~\bibnamefont
  {Holm}},\ }\href@noop {} {\emph {\bibinfo {title} {Waves with Power-Law
  Attenuation}}}\ (\bibinfo  {publisher} {Springer and ASA Press},\ \bibinfo
  {address} {Switzerland},\ \bibinfo {year} {2019})\ pp.\ \bibinfo {pages}
  {1--312}\BibitemShut {NoStop}%
\bibitem [{\citenamefont {Lasia}(2027)}]{lasia2002electrochemical}%
  \BibitemOpen
  \bibfield  {author} {\bibinfo {author} {\bibfnamefont {A.}~\bibnamefont
  {Lasia}},\ }\href@noop {} {\emph {\bibinfo {title} {Electrochemical impedance
  spectroscopy and its applications}}}\ (\bibinfo  {publisher} {Springer},\
  \bibinfo {year} {2027})\ pp.\ \bibinfo {pages} {1--376}\BibitemShut {NoStop}%
\bibitem [{\citenamefont {Orazem}\ and\ \citenamefont
  {Tribollet}(2008)}]{orazem2008electrochemical}%
  \BibitemOpen
  \bibfield  {author} {\bibinfo {author} {\bibfnamefont {M.~E.}\ \bibnamefont
  {Orazem}}\ and\ \bibinfo {author} {\bibfnamefont {B.}~\bibnamefont
  {Tribollet}},\ }\href@noop {} {\emph {\bibinfo {title} {Electrochemical
  impedance spectroscopy}}}\ (\bibinfo  {publisher} {Wiley},\ \bibinfo {year}
  {2008})\ pp.\ \bibinfo {pages} {1--525}\BibitemShut {NoStop}%
\bibitem [{\citenamefont {Hirschorn}\ \emph
  {et~al.}(2010{\natexlab{a}})\citenamefont {Hirschorn}, \citenamefont
  {Orazem}, \citenamefont {Tribollet}, \citenamefont {Vivier}, \citenamefont
  {Frateur},\ and\ \citenamefont {Musiani}}]{hirschorn2010constant}%
  \BibitemOpen
  \bibfield  {author} {\bibinfo {author} {\bibfnamefont {B.}~\bibnamefont
  {Hirschorn}}, \bibinfo {author} {\bibfnamefont {M.~E.}\ \bibnamefont
  {Orazem}}, \bibinfo {author} {\bibfnamefont {B.}~\bibnamefont {Tribollet}},
  \bibinfo {author} {\bibfnamefont {V.}~\bibnamefont {Vivier}}, \bibinfo
  {author} {\bibfnamefont {I.}~\bibnamefont {Frateur}},\ and\ \bibinfo {author}
  {\bibfnamefont {M.}~\bibnamefont {Musiani}},\ }\bibfield  {title} {\bibinfo
  {title} {Constant-phase-element behavior caused by resistivity distributions
  in films: I. theory},\ }\href@noop {} {\bibfield  {journal} {\bibinfo
  {journal} {J. Electrochem.}\ }\textbf {\bibinfo {volume} {157}},\ \bibinfo
  {pages} {C452} (\bibinfo {year} {2010}{\natexlab{a}})}\BibitemShut {NoStop}%
\bibitem [{\citenamefont {Sharma}\ \emph {et~al.}(2019)\citenamefont {Sharma},
  \citenamefont {Holm}, \citenamefont {Diaz-Real},\ and\ \citenamefont
  {M{\'e}rida}}]{sharma2019experimental}%
  \BibitemOpen
  \bibfield  {author} {\bibinfo {author} {\bibfnamefont {T.}~\bibnamefont
  {Sharma}}, \bibinfo {author} {\bibfnamefont {T.}~\bibnamefont {Holm}},
  \bibinfo {author} {\bibfnamefont {J.}~\bibnamefont {Diaz-Real}},\ and\
  \bibinfo {author} {\bibfnamefont {W.}~\bibnamefont {M{\'e}rida}},\ }\bibfield
   {title} {\bibinfo {title} {Experimental verification of pore impedance
  theory: Drilled graphite electrodes with gradually more complex pore size
  distribution},\ }\href@noop {} {\bibfield  {journal} {\bibinfo  {journal}
  {Electrochim. Acta}\ }\textbf {\bibinfo {volume} {317}},\ \bibinfo {pages}
  {528} (\bibinfo {year} {2019})}\BibitemShut {NoStop}%
\bibitem [{\citenamefont {Martinsen}\ \emph {et~al.}(1997)\citenamefont
  {Martinsen}, \citenamefont {Grimnes},\ and\ \citenamefont
  {Sveen}}]{martinsen1997dielectric}%
  \BibitemOpen
  \bibfield  {author} {\bibinfo {author} {\bibfnamefont {{\O}.~G.}\
  \bibnamefont {Martinsen}}, \bibinfo {author} {\bibfnamefont {S.}~\bibnamefont
  {Grimnes}},\ and\ \bibinfo {author} {\bibfnamefont {O.}~\bibnamefont
  {Sveen}},\ }\bibfield  {title} {\bibinfo {title} {Dielectric properties of
  some keratinised tissues. part 1: Stratum corneum and nail in situ},\
  }\href@noop {} {\bibfield  {journal} {\bibinfo  {journal} {Med. Biol. Eng.
  Comput.}\ }\textbf {\bibinfo {volume} {35}},\ \bibinfo {pages} {172}
  (\bibinfo {year} {1997})}\BibitemShut {NoStop}%
\bibitem [{\citenamefont {Zhang}\ \emph {et~al.}(2018)\citenamefont {Zhang},
  \citenamefont {Hu}, \citenamefont {Wang}, \citenamefont {Sun},\ and\
  \citenamefont {Dorrell}}]{zhang2018review}%
  \BibitemOpen
  \bibfield  {author} {\bibinfo {author} {\bibfnamefont {L.}~\bibnamefont
  {Zhang}}, \bibinfo {author} {\bibfnamefont {X.}~\bibnamefont {Hu}}, \bibinfo
  {author} {\bibfnamefont {Z.}~\bibnamefont {Wang}}, \bibinfo {author}
  {\bibfnamefont {F.}~\bibnamefont {Sun}},\ and\ \bibinfo {author}
  {\bibfnamefont {D.~G.}\ \bibnamefont {Dorrell}},\ }\bibfield  {title}
  {\bibinfo {title} {A review of supercapacitor modeling, estimation, and
  applications: A control/management perspective},\ }\href@noop {} {\bibfield
  {journal} {\bibinfo  {journal} {Renew. Sust. Energ. Rev.}\ }\textbf {\bibinfo
  {volume} {81}},\ \bibinfo {pages} {1868} (\bibinfo {year}
  {2018})}\BibitemShut {NoStop}%
\bibitem [{\citenamefont {Pandey}\ and\ \citenamefont
  {Holm}(2016)}]{pandey2016linking}%
  \BibitemOpen
  \bibfield  {author} {\bibinfo {author} {\bibfnamefont {V.}~\bibnamefont
  {Pandey}}\ and\ \bibinfo {author} {\bibfnamefont {S.}~\bibnamefont {Holm}},\
  }\bibfield  {title} {\bibinfo {title} {Linking the fractional derivative and
  the {L}omnitz creep law to non-{N}ewtonian time-varying viscosity},\
  }\href@noop {} {\bibfield  {journal} {\bibinfo  {journal} {Phys. Rev. E}\
  }\textbf {\bibinfo {volume} {94}},\ \bibinfo {pages} {032606} (\bibinfo
  {year} {2016})}\BibitemShut {NoStop}%
\bibitem [{\citenamefont {Yang}\ \emph {et~al.}(2020)\citenamefont {Yang},
  \citenamefont {Cai}, \citenamefont {Liang},\ and\ \citenamefont
  {Holm}}]{yang2020novel}%
  \BibitemOpen
  \bibfield  {author} {\bibinfo {author} {\bibfnamefont {X.}~\bibnamefont
  {Yang}}, \bibinfo {author} {\bibfnamefont {W.}~\bibnamefont {Cai}}, \bibinfo
  {author} {\bibfnamefont {Y.}~\bibnamefont {Liang}},\ and\ \bibinfo {author}
  {\bibfnamefont {S.}~\bibnamefont {Holm}},\ }\bibfield  {title} {\bibinfo
  {title} {A novel representation of time-varying viscosity with power-law and
  comparative study},\ }\href@noop {} {\bibfield  {journal} {\bibinfo
  {journal} {Int. J. Non-Lin. Mech.}\ }\textbf {\bibinfo {volume} {119}},\
  \bibinfo {pages} {103372} (\bibinfo {year} {2020})}\BibitemShut {NoStop}%
\bibitem [{\citenamefont {Buckingham}(2000)}]{buckingham2000wave}%
  \BibitemOpen
  \bibfield  {author} {\bibinfo {author} {\bibfnamefont {M.~J.}\ \bibnamefont
  {Buckingham}},\ }\bibfield  {title} {\bibinfo {title} {Wave propagation,
  stress relaxation, and grain-to-grain shearing in saturated, unconsolidated
  marine sediments},\ }\href@noop {} {\bibfield  {journal} {\bibinfo  {journal}
  {J.\ Acoust.\ Soc.\ Am.}\ }\textbf {\bibinfo {volume} {108}},\ \bibinfo
  {pages} {2796} (\bibinfo {year} {2000})}\BibitemShut {NoStop}%
\bibitem [{\citenamefont {Radwan}\ and\ \citenamefont
  {Elwakil}(2012)}]{radwan2012expression}%
  \BibitemOpen
  \bibfield  {author} {\bibinfo {author} {\bibfnamefont {A.~G.}\ \bibnamefont
  {Radwan}}\ and\ \bibinfo {author} {\bibfnamefont {A.~S.}\ \bibnamefont
  {Elwakil}},\ }\bibfield  {title} {\bibinfo {title} {An expression for the
  voltage response of a current-excited fractance device based on
  fractional-order trigonometric identities},\ }\href@noop {} {\bibfield
  {journal} {\bibinfo  {journal} {Int. J. Circ. Theor. App.}\ }\textbf
  {\bibinfo {volume} {40}},\ \bibinfo {pages} {533} (\bibinfo {year}
  {2012})}\BibitemShut {NoStop}%
\bibitem [{\citenamefont {Abdelouahab}\ \emph {et~al.}(2014)\citenamefont
  {Abdelouahab}, \citenamefont {Lozi},\ and\ \citenamefont
  {Chua}}]{abdelouahab2014memfractance}%
  \BibitemOpen
  \bibfield  {author} {\bibinfo {author} {\bibfnamefont {M.-S.}\ \bibnamefont
  {Abdelouahab}}, \bibinfo {author} {\bibfnamefont {R.}~\bibnamefont {Lozi}},\
  and\ \bibinfo {author} {\bibfnamefont {L.}~\bibnamefont {Chua}},\ }\bibfield
  {title} {\bibinfo {title} {Memfractance: a mathematical paradigm for circuit
  elements with memory},\ }\href@noop {} {\bibfield  {journal} {\bibinfo
  {journal} {Int. J. Bifurcat. Chaos}\ }\textbf {\bibinfo {volume} {24}},\
  \bibinfo {pages} {1430023} (\bibinfo {year} {2014})}\BibitemShut {NoStop}%
\bibitem [{\citenamefont {Chhabra}(2010)}]{Chhabra2010}%
  \BibitemOpen
  \bibfield  {author} {\bibinfo {author} {\bibfnamefont {R.~P.}\ \bibnamefont
  {Chhabra}},\ }\bibinfo {title} {Rheology of complex fluids}\ (\bibinfo
  {publisher} {Springer New York},\ \bibinfo {address} {New York, NY},\
  \bibinfo {year} {2010})\ Chap.\ \bibinfo {chapter} {1. Non-Newtonian Fluids:
  An Introduction}, pp.\ \bibinfo {pages} {3--34}\BibitemShut {NoStop}%
\bibitem [{\citenamefont {Biolek}\ \emph {et~al.}(2007)\citenamefont {Biolek},
  \citenamefont {Kolka},\ and\ \citenamefont {Biolkova}}]{biolek2007modeling}%
  \BibitemOpen
  \bibfield  {author} {\bibinfo {author} {\bibfnamefont {D.}~\bibnamefont
  {Biolek}}, \bibinfo {author} {\bibfnamefont {Z.}~\bibnamefont {Kolka}},\ and\
  \bibinfo {author} {\bibfnamefont {V.}~\bibnamefont {Biolkova}},\ }\bibfield
  {title} {\bibinfo {title} {Modeling time-varying storage components in
  {PS}pice},\ }in\ \href@noop {} {\emph {\bibinfo {booktitle} {Proc. Electronic
  Devices and Systems IMAPS CS International Conference EDS}}},\ Vol.\ \bibinfo
  {volume} {2007}\ (\bibinfo {organization} {Citeseer},\ \bibinfo {year}
  {2007})\ pp.\ \bibinfo {pages} {39--44}\BibitemShut {NoStop}%
\bibitem [{Spe()}]{SpectrumSoftware}%
  \BibitemOpen
  \href@noop {} {\bibinfo {title} {{Spectrum Software, Micro-Cap 12}}},\
  \bibinfo {howpublished} {\url{http://http://www.spectrum-soft.com/}},\
  \bibinfo {note} {accessed: 2020-05-18}\BibitemShut {NoStop}%
\bibitem [{\citenamefont {Johnsen}\ \emph {et~al.}(2011)\citenamefont
  {Johnsen}, \citenamefont {L{\"u}tken}, \citenamefont {Martinsen},\ and\
  \citenamefont {Grimnes}}]{johnsen2011memristive}%
  \BibitemOpen
  \bibfield  {author} {\bibinfo {author} {\bibfnamefont {G.~K.}\ \bibnamefont
  {Johnsen}}, \bibinfo {author} {\bibfnamefont {C.~A.}\ \bibnamefont
  {L{\"u}tken}}, \bibinfo {author} {\bibfnamefont {{\O}.~G.}\ \bibnamefont
  {Martinsen}},\ and\ \bibinfo {author} {\bibfnamefont {S.}~\bibnamefont
  {Grimnes}},\ }\bibfield  {title} {\bibinfo {title} {Memristive model of
  electro-osmosis in skin},\ }\href@noop {} {\bibfield  {journal} {\bibinfo
  {journal} {Phys. Rev. E}\ }\textbf {\bibinfo {volume} {83}},\ \bibinfo
  {pages} {031916} (\bibinfo {year} {2011})}\BibitemShut {NoStop}%
\bibitem [{\citenamefont {Pabst}\ \emph {et~al.}(2018)\citenamefont {Pabst},
  \citenamefont {Martinsen},\ and\ \citenamefont {Chua}}]{pabst2018non}%
  \BibitemOpen
  \bibfield  {author} {\bibinfo {author} {\bibfnamefont {O.}~\bibnamefont
  {Pabst}}, \bibinfo {author} {\bibfnamefont {{\O}.~G.}\ \bibnamefont
  {Martinsen}},\ and\ \bibinfo {author} {\bibfnamefont {L.}~\bibnamefont
  {Chua}},\ }\bibfield  {title} {\bibinfo {title} {The non-linear electrical
  properties of human skin make it a generic memristor},\ }\href@noop {}
  {\bibfield  {journal} {\bibinfo  {journal} {Sci. Rep.}\ }\textbf {\bibinfo
  {volume} {8}},\ \bibinfo {pages} {1} (\bibinfo {year} {2018})}\BibitemShut
  {NoStop}%
\bibitem [{\citenamefont {Di~Ventra}\ \emph {et~al.}(2009)\citenamefont
  {Di~Ventra}, \citenamefont {Pershin},\ and\ \citenamefont
  {Chua}}]{di2009circuit}%
  \BibitemOpen
  \bibfield  {author} {\bibinfo {author} {\bibfnamefont {M.}~\bibnamefont
  {Di~Ventra}}, \bibinfo {author} {\bibfnamefont {Y.~V.}\ \bibnamefont
  {Pershin}},\ and\ \bibinfo {author} {\bibfnamefont {L.~O.}\ \bibnamefont
  {Chua}},\ }\bibfield  {title} {\bibinfo {title} {Circuit elements with
  memory: memristors, memcapacitors, and meminductors},\ }\href@noop {}
  {\bibfield  {journal} {\bibinfo  {journal} {Proc. IEEE}\ }\textbf {\bibinfo
  {volume} {97}},\ \bibinfo {pages} {1717} (\bibinfo {year}
  {2009})}\BibitemShut {NoStop}%
\bibitem [{\citenamefont {Yamamoto}\ and\ \citenamefont
  {Yamamoto}(1981)}]{yamamoto1981non}%
  \BibitemOpen
  \bibfield  {author} {\bibinfo {author} {\bibfnamefont {T.}~\bibnamefont
  {Yamamoto}}\ and\ \bibinfo {author} {\bibfnamefont {Y.}~\bibnamefont
  {Yamamoto}},\ }\bibfield  {title} {\bibinfo {title} {Non-linear electrical
  properties of skin in the low frequency range},\ }\href@noop {} {\bibfield
  {journal} {\bibinfo  {journal} {Med. Biol. Eng. Comput.}\ }\textbf {\bibinfo
  {volume} {19}},\ \bibinfo {pages} {302} (\bibinfo {year} {1981})}\BibitemShut
  {NoStop}%
\bibitem [{\citenamefont {Kalv{\o}y}\ \emph {et~al.}(2009)\citenamefont
  {Kalv{\o}y}, \citenamefont {Frich}, \citenamefont {Grimnes}, \citenamefont
  {Martinsen}, \citenamefont {Hol},\ and\ \citenamefont
  {Stubhaug}}]{kalvoy2009impedance}%
  \BibitemOpen
  \bibfield  {author} {\bibinfo {author} {\bibfnamefont {H.}~\bibnamefont
  {Kalv{\o}y}}, \bibinfo {author} {\bibfnamefont {L.}~\bibnamefont {Frich}},
  \bibinfo {author} {\bibfnamefont {S.}~\bibnamefont {Grimnes}}, \bibinfo
  {author} {\bibfnamefont {{\O}.~G.}\ \bibnamefont {Martinsen}}, \bibinfo
  {author} {\bibfnamefont {P.~K.}\ \bibnamefont {Hol}},\ and\ \bibinfo {author}
  {\bibfnamefont {A.}~\bibnamefont {Stubhaug}},\ }\bibfield  {title} {\bibinfo
  {title} {Impedance-based tissue discrimination for needle guidance},\
  }\href@noop {} {\bibfield  {journal} {\bibinfo  {journal} {Physiol. Meas.}\
  }\textbf {\bibinfo {volume} {30}},\ \bibinfo {pages} {129} (\bibinfo {year}
  {2009})}\BibitemShut {NoStop}%
\bibitem [{\citenamefont {Brug}\ \emph {et~al.}(1984)\citenamefont {Brug},
  \citenamefont {Van Den~Eeden}, \citenamefont {Sluyters-Rehbach},\ and\
  \citenamefont {Sluyters}}]{brug1984analysis}%
  \BibitemOpen
  \bibfield  {author} {\bibinfo {author} {\bibfnamefont {G.}~\bibnamefont
  {Brug}}, \bibinfo {author} {\bibfnamefont {A.}~\bibnamefont {Van Den~Eeden}},
  \bibinfo {author} {\bibfnamefont {M.}~\bibnamefont {Sluyters-Rehbach}},\ and\
  \bibinfo {author} {\bibfnamefont {J.}~\bibnamefont {Sluyters}},\ }\bibfield
  {title} {\bibinfo {title} {The analysis of electrode impedances complicated
  by the presence of a constant phase element},\ }\href@noop {} {\bibfield
  {journal} {\bibinfo  {journal} {J. Electroanal.}\ }\textbf {\bibinfo {volume}
  {176}},\ \bibinfo {pages} {275} (\bibinfo {year} {1984})}\BibitemShut
  {NoStop}%
\bibitem [{\citenamefont {Hirschorn}\ \emph
  {et~al.}(2010{\natexlab{b}})\citenamefont {Hirschorn}, \citenamefont
  {Orazem}, \citenamefont {Tribollet}, \citenamefont {Vivier}, \citenamefont
  {Frateur},\ and\ \citenamefont {Musiani}}]{hirschorn2010determination}%
  \BibitemOpen
  \bibfield  {author} {\bibinfo {author} {\bibfnamefont {B.}~\bibnamefont
  {Hirschorn}}, \bibinfo {author} {\bibfnamefont {M.~E.}\ \bibnamefont
  {Orazem}}, \bibinfo {author} {\bibfnamefont {B.}~\bibnamefont {Tribollet}},
  \bibinfo {author} {\bibfnamefont {V.}~\bibnamefont {Vivier}}, \bibinfo
  {author} {\bibfnamefont {I.}~\bibnamefont {Frateur}},\ and\ \bibinfo {author}
  {\bibfnamefont {M.}~\bibnamefont {Musiani}},\ }\bibfield  {title} {\bibinfo
  {title} {Determination of effective capacitance and film thickness from
  constant-phase-element parameters},\ }\href@noop {} {\bibfield  {journal}
  {\bibinfo  {journal} {Electrochim. Acta}\ }\textbf {\bibinfo {volume} {55}},\
  \bibinfo {pages} {6218} (\bibinfo {year} {2010}{\natexlab{b}})}\BibitemShut
  {NoStop}%
\bibitem [{\citenamefont {Hsu}\ and\ \citenamefont
  {Mansfeld}(2001)}]{hsu2001concerning}%
  \BibitemOpen
  \bibfield  {author} {\bibinfo {author} {\bibfnamefont {C.}~\bibnamefont
  {Hsu}}\ and\ \bibinfo {author} {\bibfnamefont {F.}~\bibnamefont {Mansfeld}},\
  }\bibfield  {title} {\bibinfo {title} {Concerning the conversion of the
  constant phase element parameter y0 into a capacitance},\ }\href@noop {}
  {\bibfield  {journal} {\bibinfo  {journal} {Corrosion}\ }\textbf {\bibinfo
  {volume} {57}},\ \bibinfo {pages} {747} (\bibinfo {year} {2001})}\BibitemShut
  {NoStop}%
\bibitem [{\citenamefont {Allagui}\ \emph {et~al.}(2016)\citenamefont
  {Allagui}, \citenamefont {Freeborn}, \citenamefont {Elwakil},\ and\
  \citenamefont {Maundy}}]{allagui2016reevaluation}%
  \BibitemOpen
  \bibfield  {author} {\bibinfo {author} {\bibfnamefont {A.}~\bibnamefont
  {Allagui}}, \bibinfo {author} {\bibfnamefont {T.~J.}\ \bibnamefont
  {Freeborn}}, \bibinfo {author} {\bibfnamefont {A.~S.}\ \bibnamefont
  {Elwakil}},\ and\ \bibinfo {author} {\bibfnamefont {B.~J.}\ \bibnamefont
  {Maundy}},\ }\bibfield  {title} {\bibinfo {title} {Reevaluation of
  performance of electric double-layer capacitors from constant-current
  charge/discharge and cyclic voltammetry},\ }\href@noop {} {\bibfield
  {journal} {\bibinfo  {journal} {Sci. Rep.}\ }\textbf {\bibinfo {volume}
  {6}},\ \bibinfo {pages} {1} (\bibinfo {year} {2016})}\BibitemShut {NoStop}%
\bibitem [{\citenamefont {Gharbi}\ \emph {et~al.}(2020)\citenamefont {Gharbi},
  \citenamefont {Tran}, \citenamefont {Tribollet}, \citenamefont {Turmine},\
  and\ \citenamefont {Vivier}}]{gharbi2020revisiting}%
  \BibitemOpen
  \bibfield  {author} {\bibinfo {author} {\bibfnamefont {O.}~\bibnamefont
  {Gharbi}}, \bibinfo {author} {\bibfnamefont {M.~T.}\ \bibnamefont {Tran}},
  \bibinfo {author} {\bibfnamefont {B.}~\bibnamefont {Tribollet}}, \bibinfo
  {author} {\bibfnamefont {M.}~\bibnamefont {Turmine}},\ and\ \bibinfo {author}
  {\bibfnamefont {V.}~\bibnamefont {Vivier}},\ }\bibfield  {title} {\bibinfo
  {title} {Revisiting cyclic voltammetry and electrochemical impedance
  spectroscopy analysis for capacitance measurements},\ }\href@noop {}
  {\bibfield  {journal} {\bibinfo  {journal} {Electrochim. Acta}\ ,\ \bibinfo
  {pages} {136109}} (\bibinfo {year} {2020})}\BibitemShut {NoStop}%
\end{thebibliography}

\end{document}